\documentclass[aps,prl,preprint,groupedaddress]{revtex4-1}
\usepackage{amsmath}
\usepackage{amssymb}
\usepackage{graphicx}

\begin{document}

\title{Unitary Quantum Error Correction without Error Detection}

\author{Hiroyuki Tomita}\email[]{tomita@alice.math.kindai.ac.jp}
\affiliation{
Research Center for Quantum Computing, Interdisciplinary Graduate School of Science and Engineering, Kinki University, 
3-4-1 Kowakae, Higashi-Osaka, 577-8502, Japan}

\author{Mikio Nakahara}\email{nakahara@math.kindai.ac.jp}
\affiliation{
Research Center for Quantum Computing, Interdisciplinary Graduate School of Science and Engineering, Kinki University, 
3-4-1 Kowakae, Higashi-Osaka, 577-8502, Japan}
\affiliation{
Department of Physics, Kinki University, 
3-4-1 Kowakae, Higashi-Osaka, 577-8502, Japan}

\date{\today}

\begin{abstract}
We propose a quantum error correction without error detection.
A quantum state $\rho_0$ combined with an ancilla state $\sigma$ is encoded 
unitarily
and an error operator is applied on the encoded state.
The recovery operation then produces a tensor product state $\rho_0 \otimes
\sigma'$. The decoding operation is combined with the recovery operation and
the state $\rho_0$ is directly reproduced without referring to the code word.
A higher rank projection operator required for a conventional operator 
quantum error correction is not necessary to implement.
Encoding and the recovery operations are implemented with 
unitary operators only,
which makes quantum error correction much easier than any other proposals. 
\end{abstract}


\maketitle

A quantum system is vulnerable against disturbance from the environment.
Environmental disturbance works as a source of decoherence and it must be
suppressed to realize a working scalable quantum computer.
Among many proposals to fight against decoherence, quantum error correcting
code (QECC) is one of the most promising strategy to overcome decoherence.
There are two main approaches to QECC to date. One employs syndrome
measurements with extra ancilla qubits,\cite{9qecc,5qecc} 
while the other does not.\cite{oqec} The latter is known as the
operator quantum error correction (OQEC). In both approaches, a qubit
state $\rho_0$ together with ancilla qubit state $\sigma$
are encoded in the code space ${\mathcal C}$ as $\rho=U_E(\rho_0
\otimes \sigma)$, on which
an error operator ${\mathcal E}$ acts subsequently. 
Here $U_E$ is the encoding unitary operator.
The recovery
operator ${\mathcal R}$ is applied on the state with an error to
reproduce the initial code word $\rho$. Then a decoding operator
$U_E^{-1}$ is applied to obtain $\rho_0 \otimes \sigma$. In other words,
QECC works if it satisfies
\begin{equation}\label{eq:convqecc}
{\mathcal R}({\mathcal E}(\rho)) = \rho \quad (\rho \in {\mathcal C}).
\end{equation}
This is certainly a sufficient condition to reproduce the qubit state $\rho_0$
via recovering of the code word $\rho$.

It is the purpose of this Letter to propose a more efficient QECC. We replace
the condition (\ref{eq:convqecc}) by
\begin{equation}\label{eq:ourqecc}
{\mathcal R}({\mathcal E}(U_E(\rho_0 \otimes \sigma))) 
= \rho_0 \otimes \sigma',
\end{equation}
where $\sigma'$ is an output ancilla state which depends on
$\sigma$ and the error operator $\mathcal E$.
An essential observation is that the output state
is a tensor product of $\rho_0$ and $\sigma'$. 
It is also important to note that although we use the code space
in encoding a qubit state, it is not referred to in the
recovery process. The recovery operation here involves the
decoding process without going through the cord space.
Since the output state is a tensor product
state, we can discard the ancillas without disturbing the
qubit state $\rho_0$.

Let us examine the simplest bit-flip error channel. A qubit state $|\psi_0
\rangle = \alpha |0\rangle
+ \beta |1 \rangle$ is encoded by introducing two ancilla qubits as
\begin{equation}
U_E(|\psi_0 \rangle |00\rangle)=
\alpha |0\rangle_L + \beta |1\rangle_L \equiv |\psi \rangle,
\end{equation}
where $|0\rangle_L =|000\rangle$ and $|1 \rangle_L =|111\rangle$
are logical qubit basis vectors.
We introduce the density matrices $\rho_0 = |\psi_0 \rangle \langle \psi_0|$
and $\rho= |\psi \rangle \langle \psi|$ to denote these pure states.

Now the error operator acts on $\rho$ as
\begin{equation}\label{eq:bferror}
\Phi(\rho) = p_0 \rho + \sum_{i=1}^3 p_i X_i \rho X_i \equiv \rho',
\end{equation}
where $\sum_{i=0}^3 p_i = 1$ and
$X_i$ stands for the bit-flip operator acting on the
$i$-th qubit. ($X_1 = X \otimes I_2 \otimes I_2$, for example, where
$X=\sigma_x$ and
$I_n$ is the unit matrix of order $n$.) Here $p_i$ is the probability with 
which $X_i$ acts on $\rho$ while $p_0$ is the probability with which
$\rho$ is left intact. We formally introduce $X_0
= {I_2}^{\otimes 3}$ to denote the latter process so that
(\ref{eq:bferror}) is rewritten as
\begin{equation}\label{eq:bferror0}
\Phi(\rho) = \sum_{i=0}^3 p_i X_i \rho X_i.
\end{equation}

The conventional error recovery operation is studied in detail for
this channel in \cite{cklnote}, in which a superoperator $\Psi$ is used as
\begin{equation}
\Psi(\rho')=P_{07}\left(\sum_{i=0}^{3} X_i \rho' X_i\right) P_{07},
\end{equation}
where $P_{ij}$ is a rank-2 projection operator onto a subspace
${\rm Span}(|i\rangle, |j \rangle)$ with decimal indices
$i$ and $j$. Explicitly, it is given by
$$
P_{07} = |0\rangle_L {}_L\langle 0|+|1\rangle_L {}_L\langle 1|\nonumber\\
= {\mathrm{diag}}(1,0,0,0,0,0,0,1).
$$
The recovery operation introduced here is a superoperator,
whose physical realization is challenging compared to unitary
operations. Furthermore the rank-2 projection operator
is also difficult for physical implementation in general.

These difficulties are avoided if the recovery operation is
implemented with unitary matrices only and the
unitary matrix acting on the error state $\rho'$
outputs a tensor product state $\rho_0 \otimes \sigma'$.
We simply discard $\sigma'$ since it does not carry
any useful information.
Let us write down the density matrix $\rho'= {\mathcal E}(\rho)$
explicitly to find the recovering unitary matrix;
\begin{eqnarray}
\rho'&=&\sum_{i=0}^3 p_i~X_i~\rho~X_i \nonumber
\\
&=&\left(
\begin{array}{cccccccc}
p_0|\alpha|^2 & 0 & 0 & 0 & 0 & 0 & 0 & p_0\alpha\beta^* \\
0 & p_3|\alpha|^2 & 0 & 0 & 0 & 0 & p_3\alpha\beta^* & 0 \\
0 & 0 & p_2|\alpha|^2 & 0 & 0 & p_2\alpha\beta^* & 0 & 0 \\
0 & 0 & 0 & p_1|\beta|^2 & p_1\alpha^*\beta & 0 & 0 & 0 \\
0 & 0 & 0 & p_1\alpha\beta^* & p_1|\alpha|^2 & 0 & 0 & 0 \\
0 & 0 & p_2\alpha^*\beta & 0 & 0 & p_2|\beta|^2 & 0 & 0 \\
0 & p_3\alpha^*\beta & 0 & 0 & 0 & 0 & p_3|\beta|^2 & 0 \\
p_0\alpha^*\beta & 0 & 0 & 0 & 0 & 0 & 0 & p_0|\beta|^2
\end{array}
\right)
\end{eqnarray}
By inspecting the above matrix, we immediately notice that a
permutation $P_{(3,4)}$ of two basis vectors $|011\rangle=|3 \rangle$ 
and $|100 \rangle=|4\rangle$, followed by a
reverse ordering operation $P_{(4,5,6,7)}$ of basis vectors 
$|4\rangle, |5\rangle, |6\rangle$
and $|7\rangle$ maps $\rho'$ a tensor product form;
\begin{eqnarray}\label{eq:3qecc}
\lefteqn{R~\rho'~R^{\dagger}}\nonumber\\
&=&\left(
\begin{array}{cccccccc}
p_0|\alpha|^2 & 0 & 0 & 0 & p_0\alpha\beta^* & 0 & 0 & 0\\
0 & p_3|\alpha|^2 & 0 & 0 & 0 & p_3\alpha\beta^* & 0 & 0\\
0 & 0 & p_2|\alpha|^2 & 0 & 0 & 0 & p_2\alpha\beta^* & 0\\
0 & 0 & 0 & p_1|\alpha|^2 & 0 & 0 & 0 & p_1\alpha\beta^*\\
p_0\alpha^*\beta & 0 & 0 & 0 & p_0|\beta|^2 & 0 & 0 & 0\\
0 & p_3\alpha^*\beta & 0 & 0 & 0 & p_3|\beta|^2 & 0 & 0\\
0 & 0 & p_2\alpha^*\beta & 0 & 0 & 0 & p_2|\beta|^2 & 0\\
0 & 0 & 0 & p_1\alpha^*\beta & 0 & 0 & 0 & p_1|\beta|^2
\end{array}
\right)\nonumber
\\
&= & \left(\begin{array}{cl}
|\alpha|^2 & \alpha\beta^*\\
\alpha^*\beta & |\beta|^2
\end{array}\right)\otimes
\left(
\begin{array}{cccc}
p_0 & 0 & 0 & 0\\
0 & p_3 & 0 & 0\\
0 & 0 & p_2 & 0\\
0 & 0 & 0 & p_1
\end{array}\right),
\end{eqnarray}
where
\begin{equation}
R = P_{(4,5,6,7)} P_{(3,4)} = P_{(3,7)}P_{(4,5,6,7)}.
\end{equation}
Although the second matrix in the bottom of Eq.~(\ref{eq:3qecc}) depends on
the error operator, the first matrix reproduces the initial state
exactly. Since the recovered state is a tensor product of $\rho_0$
and an ancillary state, we can safely discard 
the ancillas without leaving any trace on the first qubit and
error correction is done. This last step corresponds to a rank-2
projection, whose physical realization is trivial in our scheme. 

\begin{figure}
\begin{center}
\includegraphics[width=10cm]{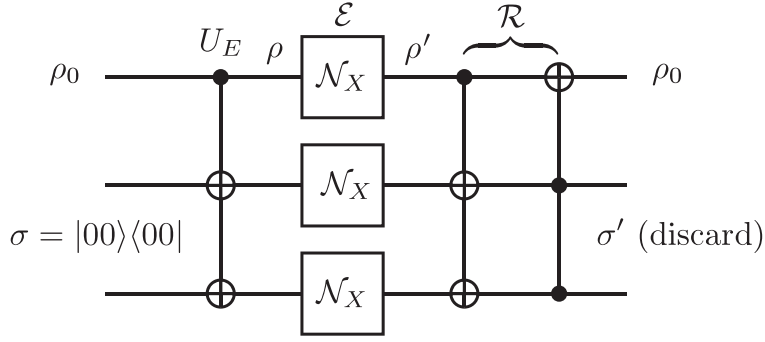}
\end{center}
\caption{3-qubit QECC for a bit-flip channel. The input state is $\rho_0
\otimes \sigma$ while the output state is $\rho_0\otimes \sigma'$.
$U_E, {\mathcal E}$ and ${\mathcal R}$ are encoding circuit, error operation,
and the ecovery operation, respectively. Needless to say, the unitary matrices
act on both sides of a density matrix (adjoint representation).
${\mathcal N}_X$ stands for the bit-flip noise.}\label{fig:3qecc}
\end{figure}

The permutation matrices $P_{(4,5,6,7)}$ and $P_{(3,7)}$ are
nothing but $[C_1 X_2 X_3]$ and $[X_1 C_2 C_3]$ gates, respectively.
Here we introduced the notation in which $C_i$ means that the $i$th
qubit works as a control bit in the gate. For example, $[C_1 X_2 X_3]$
stands for the controlled-NOT-NOT gate in conventional nomenclature,
which has been used for encoding in Eq.~(3).
In summary, encoding, error and recovery operations for this QECC are
depicted as Fig~\ref{fig:3qecc}

It is instructive to rewrite the recovery operator in a different
form to obtain a hint to find a recovery matrix $R$ for more
complicated cases. Let $\{|i \rangle\}$ be the set of basis
vectors arranged as $\{|000 \rangle, |001 \rangle, |010 \rangle,
\ldots, |110 \rangle, |111 \rangle\}$ and $\{|i'\rangle\}$ be
the set of basis vector after the permutation operation
$P_{(4,5,6,7)} P_{(3,4)}$ is applied, namely,
\begin{eqnarray}
\{|i'\rangle \}&=&
\{|000\rangle, |001\rangle,|010\rangle,|100\rangle,|111\rangle,
|110\rangle,|101\rangle,|011\rangle\}\nonumber\\
&=& \{|0\rangle_L, X_3|0\rangle_L ,X_2|0\rangle_L,
X_1|0\rangle_L,
|1\rangle_L,X_3 |1\rangle_L,X_2|1\rangle_L,
X_1|1\rangle_L\}.
\end{eqnarray}
Then the recovery operator has the matrix elements $R_{ij} = \langle i'|j 
\rangle$. Explicitly $R$ has the matrix form
\begin{equation}
R=\left( \begin{array}{c}
{}_L\langle 0|\\
{}_L\langle 0|X_3\\
{}_L\langle 0|X_2\\
{}_L\langle 0|X_1\\
{}_L\langle 1|\\
{}_L\langle 1|X_3\\
{}_L\langle 1|X_2\\
{}_L\langle 1|X_1
\end{array} \right) .
\end{equation}
Note that this recovery matrix acts on a state vector with an error
$X_i$ as
\begin{equation}
R X_i|\psi\rangle=|\psi_0\rangle \otimes |**\rangle
\end{equation}
due to the orthnormality ${}_L\langle m|X_i^T X_j |n \rangle_L 
= \delta_{ij}
\delta_{mn}$.
The ancilla state $|** \rangle $ depend on $i$, i.e., it tells us which
error operator has acted on $|\psi \rangle$.

\begin{figure}
\begin{center}
\includegraphics[width=15cm]{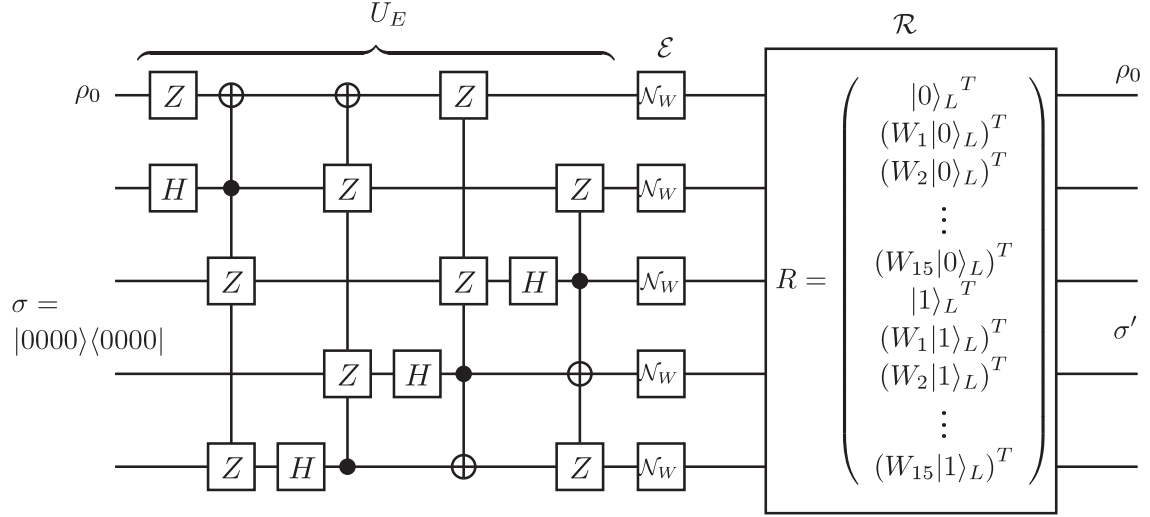}
\end{center}
\caption{5-qubit QECC. The input state is $\rho_0 \otimes |0000\rangle
\langle 0000|$ while the output state is $\rho_0\otimes \sigma'$.
$U_E, {\mathcal E}$ and ${\mathcal R}$ are encoding circuit, error operation,
and recovery operation, respectively. ${\mathcal N}_W$ stands for the
15 noise operators.}\label{fig:5qecc}
\end{figure}

Now we are ready to extend our result to more complicated QECC, 
such as Shor's 9-qubit QECC\cite{9qecc} or the DiVincenzo-Shor 5-qubit 
QECC.\cite{5qecc} 
Let us work out
the latter QECC for definiteness. Now the error operator
${\mathcal E}$ involves
$X_i, Y_i$ and $Z_i$, where $Y_i$ and $Z_i$ stand for $Y=-i \sigma_y$
and $Z=\sigma_z$, respectively, acting on the $i$th qubit.
Let $U_E$ be the encoding operator
\begin{equation}
U_E\left[(\alpha|0 \rangle + \beta 
|1\rangle)|0000\rangle\right] =\alpha|0\rangle_L + 
\beta|1\rangle_L,
\end{equation}
where the logical qubit basis vectors are
\begin{eqnarray}
|0 \rangle_L &=& \frac{1}{4}\big[|00000 \rangle 
+ |11000 \rangle +|01100\rangle +|00110\rangle +|00011\rangle +|10001\rangle
\nonumber \\
&& -|10100\rangle -|01010\rangle -|00101\rangle -|10010\rangle -|01001\rangle
\nonumber \\
& & -|11110\rangle -|01111\rangle -|10111\rangle -|11011\rangle -|11101\rangle
\big]
\label{qecc:eq:sym0}
\end{eqnarray}
and
\begin{eqnarray}
|1 \rangle_L 
&=& \frac{1}{4} \big[|11111\rangle
+ |00111\rangle+|10011\rangle +|11001\rangle +|11100\rangle +|01110\rangle
\nonumber \\
& & -|01011\rangle -|10101\rangle -|11010\rangle -|01101\rangle -|10110\rangle
\nonumber \\
& & -|00001\rangle -|10000\rangle -|01000\rangle -|00100\rangle -|00010\rangle\big].
\label{qecc:eq:sym1}
\end{eqnarray}
Let ${\mathcal E}$ be the error operator which is expressed in terms of
the operators
\begin{equation}
\{I_{32}, X_1, X_2, \ldots, X_5, Y_1, Y_2, \ldots, Y_5, Z_1, Z_2, \ldots, Z_5\}.\end{equation}
Here $X_1= X\otimes I^{\otimes 4}$ for example. It turns out to be convenient
to rename the above operators as $\{W_i\}_{0 \leq i \leq 15}$. 
For example, we have $W_0 = I_{32}, W_1= X_1, \ldots, W_{15}= Z_5$.
Suppose
\begin{equation}
\rho'={\mathcal E}(\rho) = \sum_{i=0}^{15} p_i W_i \rho W_i^T 
\end{equation}
be the error state to be recovered. We try the following recovery matrix, which
is inspired by the 3-qubit bit-flip QECC example,
\begin{equation}
R=\left(\begin{array}{c}
{|0\rangle_L}^{T}\\
{(W_1|0\rangle_L)}^{T}\\
{(W_2|0\rangle_L)}^{T}\\
\vdots\\
{(W_{15}|0\rangle_L)}^{T}\\
{|1\rangle_L}^{T}\\
{(W_1|1\rangle_L)}^{T}\\
{(W_2|1\rangle_L)}^{T}\\
\vdots\\
{(W_{15}|1\rangle_L)}^{T}
\end{array}\right).
\end{equation}
The application of $R$ on an error state ${\mathcal E}(\rho)$ outputs the 
following state
\begin{equation}
R \rho' R^T= \rho_0 \otimes \sigma',
\end{equation}
where $\rho_0 =|\psi_0 \rangle \langle \psi_0|$ as before and
\begin{equation}
\sigma'={\rm diag}(p_0, p_1, \ldots, p_{14}, p_{15}).
\end{equation}
A quantum circuit which implements the five-qubit QECC is shown in 
Fig.~\ref{fig:5qecc}, in which the encoding circuit is taken from 
\cite{nakahara_ohmi}.

The matrix $R$ is orthogonal and can be implemented with elementary
quantum gates such as one-qubit gates and CNOT gates in principle.
Nonetheless, it is not a simple permutation gate any more due to the
complicated structure of the logical qubit states $|0\rangle_L$ 
and $|1 \rangle_L$ and its implementation must be challenging.
Let us look at Fig.~\ref{fig:3qecc} to find a hint to overcome
this problem. The recovery
operation is made of the inverse encoding circuit $[C_1 X_1 X_2]$
and a permutation of basis vectors given by
the controlled-controlled NOT gate $[X_1 C_2 C_3]$. 
We also tried the inverse encoding
circuit ${U_E}^{-1}$ in our 5-qubit. Then it turned out that the
first qubit state does not agree with $|\psi_0 \rangle$ exactly for
some errors
and we need additional bit-flips and/or phase-flips to correct
this. Circuit implementation of this recovery operation will be
reported elsewhere.

\begin{figure}
\begin{center}
\includegraphics[width=13cm]{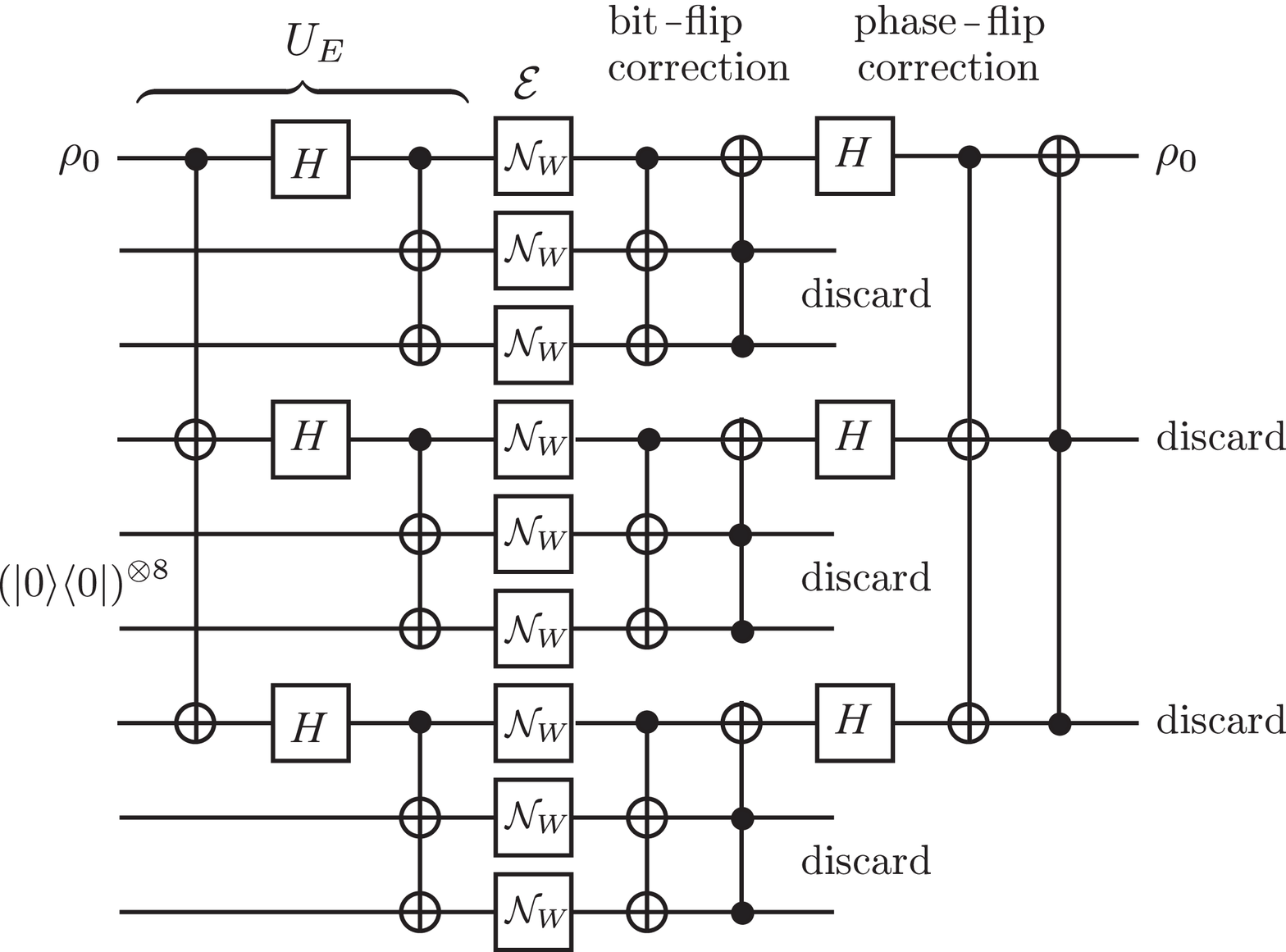}
\end{center}
\caption{}\label{fig:9qecc}
\end{figure}

Shor's 9-qubit QECC is also implemented with a unitary recovery matrix.
It is a trivial generalization of the 3-qubit bit-flip QECC and we simply
give the circuit for this case in Fig.~\ref{fig:9qecc} without giving the 
lengthy details, which will be reported elsewhere.

In summary, we proposed an efficient implementation of QECC,
whose recovery process involves unitary operations only.
No syndrome readouts nor higher-rank projection operators are
required. This makes physical implementation of QECC considerably
easier. We have demonstrated our proposal with 3-qubit bit-flip
QECC, DiVincenzo-Shor's 5-qubit QECC and Shor's 9-qubit QECC.
Details of our QECC, involving efficient decomposition of the
recovery operation into elementary gates, are in preparation and will be 
reported elsewhere.

\begin{acknowledgments}

MN would like to thank Chi-Kwong Li for sending us
\cite{cklnote} prior to publication.
A part of this research is supported by
``Open Research Center''
Project for Private Universities: Matching fund subsidy from
MEXT (Ministry of Education, Culture, Sports, Science and
Technology).

\end{acknowledgments}


\end{document}